\documentclass[runningheads]{llncs}
\usepackage[T1]{fontenc}
\usepackage{graphicx}
\usepackage{booktabs}
\usepackage[table,xcdraw]{xcolor}
\usepackage[table,xcdraw]{xcolor}

\usepackage{subcaption}
\usepackage{hyperref}
\usepackage{framed}
\usepackage{xcolor}
\usepackage{array}
\usepackage{multirow}
\usepackage{comment}
\usepackage{amsmath}

\usepackage{algorithm}
\usepackage{caption}

\begin{document}
\title{How good are LLMs at Retrieving Documents in a Specific Domain?}
%
%
\author{Nafis Tanveer Islam\inst{1,2}\and
Zhiming Zhao\inst{1,3, *}}

%
\authorrunning{N. Islam et al.}
%

\institute{
MultiScale Networked Systems (MNS) \and
University of Amsterdam, Netherlands, 
\email{n.t.islam@uva.nl} \and
University of Amsterdam, Netherlands,
\email{z.zhao@uva.nl}\\
}
\maketitle              


\newcommand\blfootnote[1]{%
  \begingroup
  \renewcommand\thefootnote{}\footnote{#1}%
  \addtocounter{footnote}{-1}%
  \endgroup
}

\begin{abstract}

Classical search engines using indexing methods in data infrastructures primarily allow keyword-based queries to retrieve content. While these indexing-based methods are highly scalable and efficient, due to a lack of an appropriate evaluation dataset and a limited understanding of semantics, they often fail to capture the user's intent and generate incomplete responses during evaluation. This problem also extends to domain-specific search systems that utilize a Knowledge Base (KB) to access data from various research infrastructures. Research infrastructures (RIs) from the environmental and earth science domain, which encompass the study of ecosystems, biodiversity, oceanography, and climate change, generate, share, and reuse large volumes of data. While there are attempts to provide a centralized search service using Elasticsearch as a knowledge base, they also face similar challenges in understanding queries with multiple intents. To address these challenges, we proposed an automated method to curate a domain-specific evaluation dataset to analyze the capability of a search system. Furthermore, we incorporate the Retrieval of Augmented Generation (RAG), powered by Large Language Models (LLMs), for high-quality retrieval of environmental domain data using natural language queries. Our quantitative and qualitative analysis of the evaluation dataset shows that LLM-based systems for information retrieval return results with higher precision when understanding queries with multiple intents, compared to Elasticsearch-based systems.
\begingroup
\renewcommand\thefootnote{}%
\footnotetext{* Corresponding Author}%
\addtocounter{footnote}{-1}%
\endgroup

\end{abstract}
\section{Introduction}

Classical search engines \cite{halavais2017search}, such as Google or academic search engines like Google Scholar, operate primarily on keyword-based retrieval mechanisms. They index vast amounts of textual or image data and return results based on exact or partial keyword matches, often ranked by measures such as term frequency \cite{das2023comparative}, PageRank \cite{brin1998anatomy}, or other heuristics. While highly efficient and scalable, these systems typically lack deep semantic understanding of the queries. A search query like \textit{"RAG papers before/after 2024"} on platforms such as Google Scholar often fails to capture the user’s true intent due to the limitations of keyword-based search systems. Instead of interpreting "before/after 2024" as a temporal filter on publication dates, the system treats the entire string as a literal phrase, retrieving documents that contain the words "before" or "after" alongside "2024"—often resulting in papers published in 2024, not those published before or after that year. Similarly domain specific search engines for research infrastructures faces similar issues due to the existence of similar underlying problem of using index based knowledge base like elasticsearch.

A knowledge base \cite{alkhamissi2022review} is a centralized repository that stores general information, facts, and rules about a specific domain or domains in the form of textual data, relational database and more. This structured information allows classical or dialogue-based search systems to perform tasks such as question-answering, reasoning, problem-solving, and decision-making by accessing and utilizing the stored knowledge. ENVRY-FAIR \cite{9041704} exemplifies a knowledge base \cite{envri-fair} that offers researchers access to domain specific data for climate change research and mitigation strategies for environmental research infrastructures. However, usage of these semi-structured knowledge-bases (SKBs) \cite{wu2024stark} supporting text-based queries on unstructured textual knowledge \cite{karpukhin2020dense}\cite{lee2019latent}, SQL based \cite{yu-etal-2018-spider}\cite{zhong2017seq2sql} or knowledge graph based \cite{he2024g} solutions are often hampered by lexical gap problems \cite{berger2000bridging}. This restricts them from generating results that fully comply with the user intent. Moreover, classical search engines built upon a knowledge-base like Elasticsearch \cite{pmlrag_2024} also face limitations such as the lack of contextual understanding and an inability to support conversational or dialogue-based queries, which enables the researcher or end user to dig deeper into a problem.

However, evaluating these knowledge bases is highly challenging, especially when they are domain-specific. This challenge primarily boils down to the lack of proper evaluation datasets because the number of end users in domain-specific research infrastructures is extremely low. In these cases, the systems are queried manually to check the quality of the response from the classical search system; an automated method to determine the quality of similar systems is not available due to the lack of appropriate evaluation data. Furthermore, to solidify our argument regarding the challenges of classical search engines, we briefly analyzed the ENVRI-FAIR \cite{envri-fair} project, which aims to centralize content from environmental Research Infrastructures (RIs) into a unified access portal. We used some of the evaluation dataset to analyze ENVRI-FAIR, a classical search system which we discuss in detail in Section \ref{sec:experiment}. For example, queries such as “North Ferriby from 1985 to 1989” in Table \ref{tab:motivation} fail to retrieve or rank any relevant datasets within the top 20 results in traditional Elasticsearch-based systems. However, the content was retrieved easily for more straightforward queries like “North Ferriby” highlighting severe gaps in the retrieval mechanism when the temporal requirement was appended to the query. These elastic search-based knowledge systems typically rely on keyword matching and struggle to interpret multi-intent queries that combine spatial and temporal constraints. This issue is further evident in other mixed-type queries, such as “Sabrina Arnold 2024-06-27” or “Brixham laboratories microbe biomass and diversity,” where relevant results were either ranked very low or not retrieved within the top 20 results. 


To address these challenges, we propose a formalized method for creating an evaluation dataset that measures search quality. Moreover, we propose a Retrieval-Augmented Generation (RAG)-based method tailored for the environmental and earth science domain for a dialogue-based information retrieval. Our process leverages a fine-tuned Large Language Model (LLM) for contextual query understanding and semantic retrieval using RAG from a vector database for knowledge storage. The fine-tuning enables the model to better understand domain-specific terminology and user intent patterns. Finally, to verify the responses generated by our system, we utilized our evaluation dataset preparation method to curate a domain-specific evaluation dataset, allowing us to analyze our system both quantitatively and qualitatively.

\begin{table}[t]
\centering
\caption{A primary analysis of how the platform ENVRI-FAIR fails to understand queries with multiple intents. The \textit{Keywords} are the search queries, and \textit{Type} specifies whether the query has single or mixed intent. \textit{Correctness} Defines whether a correct result stays in the top 20 results, and \textit{Order} shows the order of the proper response in the top 20 results.}
\begin{tabular}{@{}llll@{}}

\toprule
\multicolumn{1}{l}{\textbf{Keywords}}                                                         & \multicolumn{1}{l}{\textbf{Type}} & \multicolumn{1}{l}{\textbf{Correctness}} & \multicolumn{1}{l}{\textbf{Order}}     \\ \midrule
\begin{tabular}[l]{@{}l@{}}Quality-controlled \\ data on molar fraction\end{tabular}          & Single                          & Yes                                      & 1                                       \\ \midrule
\begin{tabular}[c]{@{}l@{}}Sabrina Arnold \\ 2024-06-27\end{tabular}                          & Mixed                             & No                                       & 12                                      \\ \midrule
\begin{tabular}[c]{@{}l@{}}Lise Lotte ICOS \\ ATC CO2 Release\end{tabular}                    & Mixed                             & Yes                                  & 3                                       \\ \midrule
\begin{tabular}[c]{@{}l@{}}Brixham laboratories microbe \\ biomass and diversity\end{tabular} & Mixed                        & No                                       & 17                                      \\ \midrule
\begin{tabular}[c]{@{}l@{}}microbe biomass and diversity\\  Brixham laboratories\end{tabular} & Mixex                        & No                                       & Not found in top 20                     \\ \midrule
\begin{tabular}[c]{@{}l@{}}Insect and earthworm \\ AstraZeneca taxonomy\end{tabular}          & Mixed                             & Yes                                      & 2                       \\ \midrule
North Ferriby                                                                                 & Single                          & Yes                                       & {1} \\ \midrule
North Ferriby From 1985 to 1989                                                               & Mixed                             & No                                       & Not found in top 20                        \\ \bottomrule
\end{tabular}
\label{tab:motivation}
\end{table}

In summary, the contributions of this paper are:

\begin{itemize}

    \item We propose a method to create an evaluation dataset for the environment and earth science domain, enabling the automatic evaluation of search results from user queries.
    
    \item We propose a RAG-based framework with a knowledge base for the environmental and earth science domain, dubbed \texttt{EnvKB} (Environment Knowledge Base), which incorporates a fine-tuned LLM to retrieve information from dialogue-based queries.


    \item We compare the capability of our proposed method with the results from an Elasticsearch-based knowledge base, ENVRI-FAIR, using a curated evaluation dataset and compare them quantitatively and qualitatively.
    
\end{itemize}

\section{Related Work}





LLMs have been used to augment search systems by enhancing queries, summarizing content, optimizing indexing, and improving result ranking \cite{liu2024information}. These methods operate within search engine frameworks, leveraging LLMs for language understanding and processing while requiring minimal computational power, primarily focusing on query manipulation. Ayoub et al. \cite{ayoub2024case} use prompt engineering to paraphrase or generate query passages. Chen et al. \cite{chen2024large} explore tailoring search experiences to user intent, and Zhou and Li \cite{zhou2024understanding} examine the shift in user behavior from traditional search engines to generative AI systems. Generative search engines have also been applied to complex knowledge tasks \cite{suri2024use}, with Li et al. \cite{li2024survey} providing an overview of their evolution. Further studies by Thomas et al. \cite{thomas2024large} address challenges like scalability and user alignment, highlighting the potential and limitations of LLM-driven retrieval systems.

The integration of LLMs into search and retrieval systems has revolutionized information retrieval, enabling more contextual and personalized search experiences. Liu et al. \cite{liu2024information} demonstrate how LLMs enhance sparse retrieval methods by refining relevance estimation and query understanding. Wang et al. \cite{wang2024large} reimagine the search stack with a LLM integrating generative capabilities for advanced query handling, while Salemi el al. \cite{salemi2024towards} optimize cross-model ranking for diverse queries. 

A dialogue-based search or question-answering-based information retrieval system \cite{xiong2024interactive}\cite{alinejad2024evaluating} is an interactive information retrieval paradigm where users engage in a multi-turn conversation with the search system to refine their queries and explore relevant content iteratively. Unlike traditional keyword-based or static query systems, dialogue-based search systems utilize natural language understanding and context retention to interpret user intent and generate results accordingly. Referring to Table \ref{tab:motivation}, queries like \textit{“North Ferriby from 1985 to 1989”} or \textit{“Brixham laboratories microbe biomass and diversity”} were either \textit{not found at all} or ranked very low in traditional search. A dialogue-based system could improve this outcome by inherently understanding the meaning by extracting and analyzing contents from a knowledge base.

However, to the best of our knowledge, none of the previous work focused on using an LLM-based knowledge base for the environment and earth science domain with retrieval using RAG. 



\section{Methodology}

\begin{figure*}[ht]
    \centering
    \includegraphics[width=1.0\linewidth]{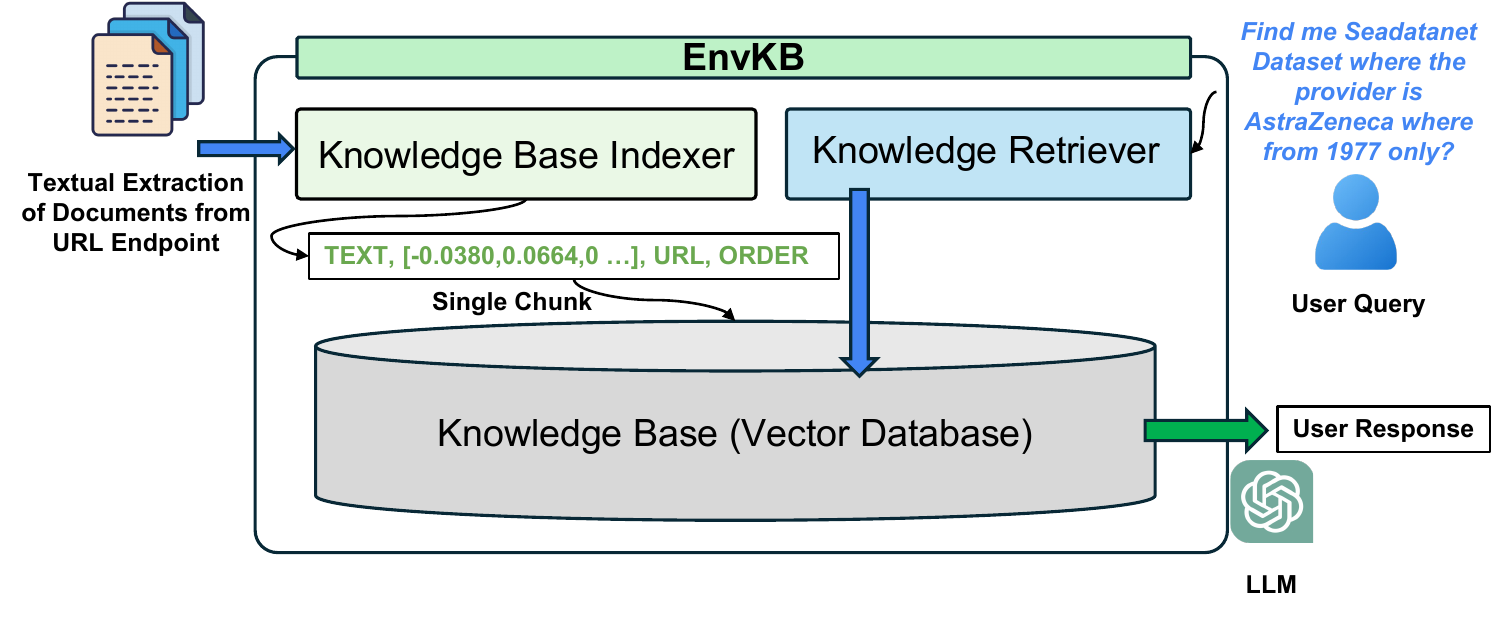}
    \caption{Overall Architecture of our Proposed System}
    \label{fig:arch}
\end{figure*}

To address the challenges of classical search engines as demonstrated through the examples in Table \ref{tab:motivation}, we propose a dialogue-based knowledge retrieval system \texttt{EnvKB}. Figure \ref{fig:arch} shows the overall architecture of our proposed system. We divided our system into three primary components namely i) Knowledge Base Indexer, ii) Knowledge Base Retriever and iii) Knowledge Base (Vector Database).

\paragraph{\textbf{Knowledge Base Indexer.}}
The goal of the Knowledge-Base Indexer of an LLM-based system is to extract documents from the URL endpoints of the environment domain. The contents of these URL endpoints should be searchable by \texttt{EnvKB}. However, these endpoints can be in different formats, including HTML, CSV, or JSON. The goal of the knowledge base indexer is to convert these formats into a text-based format for the LLM to understand. However, LLMs have a limit on the number of tokens they can handle. Therefore, the knowledge base indexer also chunks the document into multiple shards if the token length exceeds a certain threshold and keeps track of the order of the shards so they can be concatenated if needed and map them with their corresponding URL. Finally, using a trained LLM, the knowledge base indexer generates the embedding vector for each of the chunks as depicted as the \textit{Single Chunk} in Figure \ref{fig:arch}.

\paragraph{\textbf{Knowledge Retriever.}} The goal of the Knowledge Retriever is to understand the intent of the query, gather related responses from the knowledge base, and use the LLM to generate an appropriate response. This structured information allows classical or dialogue-based search systems to perform tasks such as question-answering, reasoning, problem-solving, and decision-making by accessing and utilizing the stored knowledge.
However, without training the LLM on the environmental domain, they face a problem called hallucination, where they tend to answer a question that does not make any sense, misses the context, or is borderline incorrect. Therefore, we also train the LLM with appropriate environmental domain data before using it for generating responses. Furthermore, we propose using Retrieval of Augmented Generation (RAG) \cite{lewis2020retrieval}, to retrieve content from the knowledge base.

\paragraph{\textbf{Knowledge Base.}}  A knowledge base \cite{alkhamissi2022review} is a centralized repository that stores general information, facts, and rules about a specific domain or domains in the form of textual data, relational database, and more. The knowledge base sits at the center of our proposed system, where the knowledge is stored and also retrieved from. Knowledge can be stored in various systems, including relational databases like SQL or stored in textual format, or semi-structured databases like Elasticsearch. While these methods are effective, it is challenging to search for the correct content from them as they only store textual content. While some works \cite{farshidi2022adaptable} have tried to store similar texts in the knowledge base to expand the search criteria, they used a basic Natural Language Processing (NLP) technique. Although this is the right direction, generating similar texts is an exhaustive task when the model is not trained on the domain. Therefore, we propose to use a trained LLM on domain-specific data and generate the embedding for each chunk and then store the combination of \textit{[TEXT, EMBEDDING, URL, ORDER]} in a vector database, which we use as a knowledge base.

\section{Experimental Results}
\label{sec:experiment}

In this section, we describe and discuss our entire experimental process, starting with curating the evaluation dataset, evaluation metrics, training, embedding generation, and finally, the quantitative and qualitative results using case studies. Our code is available here \footnote{\url{https://github.com/QCDIS/Retrieval_Paper/tree/main}}.

\subsection{Evaluation Dataset Preparation}

We create an evaluation dataset to evaluate the effectiveness of our proposed system \texttt{EnvKB} on retrieving relevant content aligned with the query. Our hypothesis on developing an evaluation dataset is based on the idea that the contents we store in the vector database or the contents that are already stored in the Elasticsearch database of ENVRI-FAIR can be extracted using the exact or similar keywords from the URL endpoints. Since we were able to find the contents that ENVRI-FAIR indexed, we used those URLs to create our evaluation dataset.

\begin{table}[t]
\centering
\caption{A few snapshots of our evaluation dataset}
\begin{tabular}{@{}lll@{}}
\toprule
\textbf{Key Value}                                                                                                            & \textbf{Type}               & \textbf{URL}                                                                                                                                                                                                      \\ \midrule
\multicolumn{1}{l|}{\begin{tabular}[c]{@{}l@{}}British Oceanographic\\  Data Centre\end{tabular}}                             & \multicolumn{1}{l|}{Single} & \url{https://edmed.seadatanet.org/report/854/}                                                                                                                                                                          \\ \midrule
\multicolumn{1}{l|}{Northern North Sea}                                                                                       & \multicolumn{1}{l|}{Single} & \url{https://edmed.seadatanet.org/report/851/}                                                                                                                                                                          \\ \midrule
\multicolumn{1}{l|}{Total Oil Marine Plc}                                                                                     & \multicolumn{1}{l|}{Single} & \begin{tabular}[c]{@{}l@{}}\url{https://edmed.seadatanet.org/report/850/}\\ \url{https://edmed.seadatanet.org/report/851/}\\ \url{/https://edmed.seadatanet.org/report/852/}\end{tabular}                                            \\ \midrule
\multicolumn{1}{l|}{\begin{tabular}[c]{@{}l@{}}AstraZeneca, \\ Brixham Environmental\\  Laboratory\end{tabular}}              & \multicolumn{1}{l|}{Mixed}  & \begin{tabular}[c]{@{}l@{}}https://edmed.seadatanet.org/report/858/\\ \url{https://edmed.seadatanet.org/report/865/}\\ \url{https://edmed.seadatanet.org/report/866/}\end{tabular}                                            \\ \midrule
\multicolumn{1}{l|}{\begin{tabular}[c]{@{}l@{}}Manual biota samplers,\\ gas chromatograph \\ mass spectrometers\end{tabular}} & \multicolumn{1}{l|}{Mixed}  & \begin{tabular}[c]{@{}l@{}}\url{https://edmed.seadatanet.org/report/894/}\\ \url{https://edmed.seadatanet.org/report/894/}\end{tabular}                                                                                       \\ \midrule
\multicolumn{1}{l|}{\begin{tabular}[c]{@{}l@{}}Scottish Environment \\ Protection Agency, \\ Stirling Office\end{tabular}}    & \multicolumn{1}{l|}{Mixed}  & \begin{tabular}[c]{@{}l@{}}\url{https://edmed.seadatanet.org/report/893/}\\  \url{https://edmed.seadatanet.org/report/887/}\\ \url{https://edmed.seadatanet.org/report/886/}\end{tabular} \\ \bottomrule
\end{tabular}
\label{tab:sample_dataset}
\end{table}

We begin by selecting a representative set of 1000 endpoints provided by ENVRI-FAIR \cite{pmlrag_2024} where the endpoints were provided at the GitHub repository \footnote{\url{https://github.com/QCDIS/kb-indexer}}. These endpoints are URLs pointing to individual content records (e.g., metadata pages, data summaries). Since these endpoints have structured format, we selected keywords like \textit{Data set name}, \textit{Country}, \textit{Geographical area}, \textit{Summary}, and more and extracted the key value pairs. A sample endpoint is exemplified here \footnote{\url{https://edmed.seadatanet.org/report/70/}}. We randomly select specific key values from these endpoints to create queries with single intent. Furthermore, some key value pair may overlap in multiple endpoint URL. Therefore we combine all the URLs for that specific key value pairs. Thus, if the system returns atleast one of the URLs for a key value, we consider them as the correct response. We put these key values in our system and the ENVRI-FAIR system and compare the results. Furthermore, we also mix different key values from the same URL endpoint to simulate mixed intent. In total we have collected 1500 samples for evaluation. Out of them 1000 are queries with single intent, and 500 are queries with multiple intent. Table \ref{tab:sample_dataset} shows some sample example of our evaluation dataset.

\subsection{Evaluation Metrics}
To measure the relevancy of the returned responses we used the evaluation metrics $Hits@K$ and $BERTScore$.

\paragraph{\textbf{Hits@K:}} Hits@K measures the proportion of queries for which at least one relevant result appears in the top-$K$ (k=10) retrieved documents. A query is considered a hit if the target document is found within the top-$K$ positions. Equation~\ref{eqn:1} defines the metric.

\begin{equation}
\label{eqn:1}
    Hits@K = \frac{\text{Number of queries with relevant results in top } K}{\text{Total number of queries}}
\end{equation}

\paragraph{\textbf{BERTScore:}} To capture semantic similarity beyond exact matches with URL, we use BERTScore \cite{zhang2019bertscore}, which computes token-level similarity between queries and retrieved responses using contextual embeddings from BERT \cite{devlin-etal-2019-bert}. For each query, we evaluate the first ranked result and average BERTScore across all queries to assess overall semantic alignment.

\subsection{Model Training and Embedding Generation}
In this work, we pre-train different models using a Sequence-to-Sequence \cite{sutskever2014sequence} language modeling architecture in order to ensure the model has the knowledge of the environmental and earth science domain. We select LLaMa 3.1 as our base model and compare our proposed system with elasticsearch based system, ENVRI-FAIR and other LLMs. For LLM comparison, we used LLaMa 2.7, Phi-3 and Phi-2 and Mistral. By leveraging an encoder-decoder framework, the model is capable of encoding complex, domain specific multi-intent input queries and generating contextually grounded outputs in natural language. We adopt the Sequence-to-Sequence task type within the PEFT (Parameter-Efficient Fine-Tuning) \cite{xu2023parameter} configuration to enable LoRA-based \cite{hu2021lora} adaptation, allowing for efficient fine-tuning of large models. This method of training allows the model to not only adopt to new data but also remember its previous training data as well. To collect data for model training we used a tool \textit{Crawl4ai} to extract the texts from the URL endpoints we mentioned earlier. We trained the model for 5 epochs with a maximum token length of 2048, and our learning rate was 2e$\wedge$ 5 with a batch size of 2 for our 3.1B parameter LLaMa model. We also use a beam size of 4 for the generation task and a temperature value of 0.5 for optimal performance. We used 1 NVIDIA A100 GPU with 40 GB of memory.

After training the model, we use the text extracted using  \textit{Crawl4ai} and chunk each document into tokens of size 250 for optimal results to generate embedding. Then, we store the embedding, the text, the chunked order, and the corresponding URL in a vector database. For retrieval, initially, the query is converted to an embedding, and we run a cosine similarity of the query embedding with the embeddings stored in the knowledge base and retrieve the top 10 results with the highest cosine similarity.


\begin{table}[t]
\centering
\caption{A comparisonal analysis on the performance of our proposed LLM with Elasticsearch based system ENVRI-FAIR and other LLMs}
\begin{tabular}{@{}l@{\hspace{1em}}l@{\hspace{1em}}l@{\hspace{2em}}l@{\hspace{1em}}l@{}}
\toprule
                 & \multicolumn{2}{c@{\hspace{3em}}}{\textbf{Single Query}} & \multicolumn{2}{c}{\textbf{Mixed Query}} \\ \midrule
                 & \textbf{Hits@10}        & \textbf{BERTScore}        & \textbf{Hits@10}        & \textbf{BERTScore}       \\ \midrule
\textbf{Elasticsearch}       & \textbf{0.96}          & 0.89             & 0.37          & 0.45            \\
\textbf{LLaMa 3.1}        & \textbf{0.96}          & \textbf{0.92}             & \textbf{0.84}          & \textbf{0.90}            \\
\textbf{LLaMa 2.7}        & 0.91          & 0.91             & 0.82          & 0.88            \\
\textbf{Mistral}        & 0.87          & 0.84             & 0.82          & 0.81            \\
\textbf{Phi-3}            & 0.91          & 0.88             & 0.77          & 0.85            \\
\textbf{Phi-2}            & 0.84          & 0.86             & 0.78          & 0.84            \\ \bottomrule
\end{tabular}
\label{tab:primary_comparison}
\end{table}

\subsection{Experimental Analysis}

The evaluation results presented in Table \ref{tab:primary_comparison} highlight the comparative performance of elasticsearch and LLM-based search systems using both Hits@10 and BERTScore across single and mixed query scenarios. Elasticsearch performs strongly on single queries (Hits@10 = 0.96), indicating its effectiveness when queries closely match indexed content. However, it significantly under-performs on mixed queries (Hits@10 = 0.37, BERTScore = 0.45), reflecting its limitations in handling multi-intent or semantically complex inputs. In contrast, all LLM-based models show improved performance in the mixed query setting, with LLaMa 3.1 achieving the highest Hits@10 (0.96) and BERTScore (0.92), suggesting a better capability to interpret and retrieve results based on semantic context. While there is a slight trade-off in single query performance compared to the elastic search-based system ENVRI-FAIR, the LLM models offer a more balanced and context-aware retrieval strategy, particularly evident in their superior handling of disjoint or compositional queries.

\begin{table}[t]
\centering
\caption{Performance comparison between base and trained LLMs for single and multiple query tasks}\begin{tabular}{@{}l@{\hspace{1.5em}}l@{\hspace{1.5em}}l@{\hspace{2em}}l@{\hspace{1.5em}}l@{}}
\toprule
\multicolumn{3}{c@{\hspace{2em}}}{\textbf{Single Query}}     & \multicolumn{2}{c}{\textbf{Multiple Query}} \\ \midrule
\textbf{LLM Name}   & \textbf{Base LLM} & \textbf{Trained LLM} & \textbf{Base LLM}   & \textbf{Trained LLM}  \\ \midrule
\textbf{LLaMa 3.1}  & 0.89              & 0.93                 & 0.84                & 0.88                  \\
\textbf{LLaMa 2.7}  & 0.90              & 0.94                 & 0.81                & 0.86                  \\
\textbf{Mistral}    & 0.88              & 0.92                 & 0.79                & 0.83                  \\
\textbf{Phi-3}      & 0.86              & 0.90                 & 0.75                & 0.80                  \\
\textbf{Phi-2}      & 0.83              & 0.87                 & 0.76                & 0.82                  \\ \bottomrule
\end{tabular}
\label{tab:training_comparison}
\end{table}

The results in Table \ref{tab:training_comparison} demonstrate the benefits of fine-tuning base LLMs for domain-specific search tasks, as shown across both single and multiple query scenarios. In every case, the trained models outperform their base counterparts on BERTScore, reflecting improvements in both understanding and retrieving relevant content. For instance, LLaMa 3.1 shows a notable increase of BERTScore from 0.89 to 0.93 in single query performance and from 0.84 to 0.88 in multiple queries, suggesting that training enhances the model’s sensitivity to both simple and compound search intents. This trend holds across all models, with Phi-3 improving from 0.83 to 0.87 in single queries and from 0.76 to 0.82 in multiple queries. These gains underline how fine-tuning helps models internalize domain-specific patterns, vocabulary, and reasoning strategies, enabling more accurate alignment between user queries and indexed content.

\subsection{Case Studies}
\begin{figure}[t]
    \centering
    \begin{subfigure}[b]{0.49\linewidth} 
        \includegraphics[width=\linewidth]{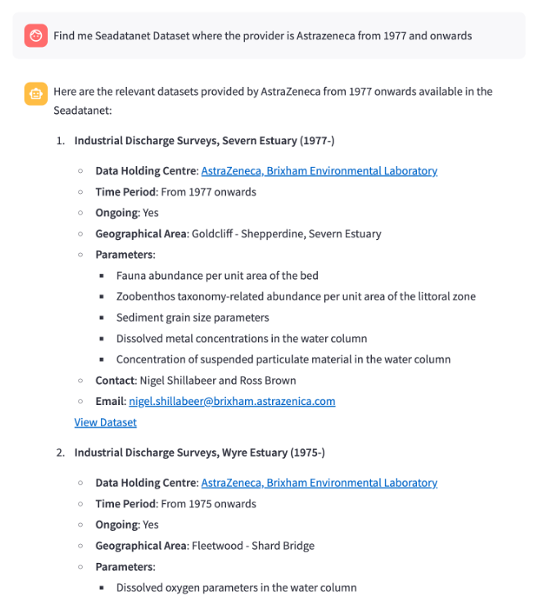}
        \caption{Case Study 1}
    \label{fig:cs1}
    \end{subfigure}
    \hfill
    \begin{subfigure}[b]{0.49\linewidth}
        \includegraphics[width=\linewidth]{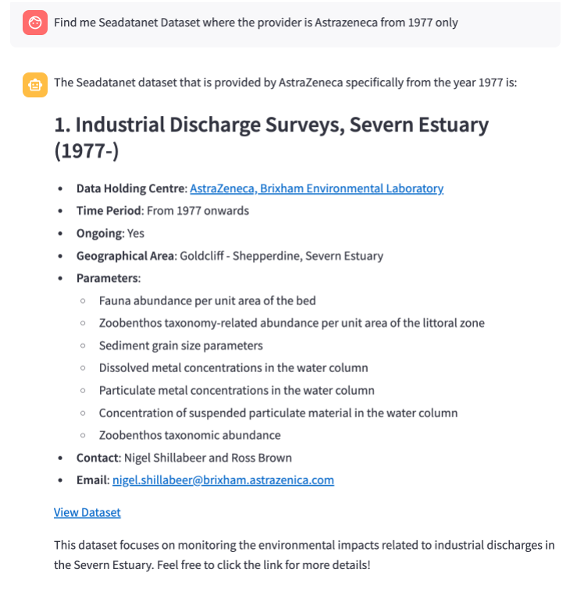}
        \caption{Case Study 2}
    \label{fig:cs2}
    \end{subfigure}
\caption{A case study of our Current System.}
\label{fig:case_studies}
\end{figure}

In this section, we will demonstrate with actual examples how our system understands queries with multiple intent. In \textbf{Case Study 1} from Figure \ref{fig:cs1} shows a query to our system "Find me Seadatanet Dataset where the provider is AstraZeneca where from 1977 and onwards". The system returned two responses for this query, the first from 1977. However, the 2nd result is from 1975. Our model is aware of this fact, mentioning that the period is From 1975 onwards. In \textbf{Case Study 2} in Figure \ref{fig:cs2}, the intent is changed slightly with the query "Find me Seadatanet Dataset where the provider is AstraZeneca where from 1977 only". In this query, we requested the system to respond with data only from 1977. From the response, we can see that the system only responded with one result from 1977. These case studies prove that our model can capture multiple intents and produce results accordingly.

\section{Conclusion}

This work examines how search in the environment and earth science domains can be enhanced using large language models (LLMs) over traditional search systems. Initially, we demonstrated a method to curate a dataset for evaluation. Then, we proposed a learning-based method to train an LLM with domain-specific knowledge, utilizing a vector database. Our comparison analysis shows that our dialogue-based system \texttt{EnvKB} can demonstrate better search results than a classical search system, ENVRI-FAIR, using Elasticsearch. Furthermore, we demonstrated the benefits of pre-training in generating more effective embeddings, which ultimately led to improved search results quality. Finally, we presented two case studies that highlight the effectiveness of our system in the real world. In our future work, we aim to demonstrate how our knowledge-based system can be scaled for various domains with continually updated and ever higher amounts of data.


%
%
%
\section*{Acknowledgment}
This research was made possible through funding from the Dutch Research Council (NWO) Large-Scale Research Infrastructures (LSRI) programme for the LTER-LIFE (http://www.lter-life.nl) infrastructure (grant 184.036.014). The work is also partially funded by several European Union projects: ENVRI-Hub Next (101131141), EVERSE (101129744), BlueCloud-2026 (101094227), OSCARS (101129751). 
\bibliographystyle{splncs04}
\bibliography{LaTeX2e+Proceedings+Templates+download/bibliography}
%





\end{document}